\documentclass[oldversion]{aa}
\usepackage{amsmath,amssymb,amscd}
\usepackage{graphicx}
\usepackage{natbib}
\bibpunct{(}{)}{;}{a}{}{,}
\usepackage{revsymb}	%JWL
\usepackage{amsmath}	%JWL
\usepackage[usenames]{color}	%JWL [pour command \JWL]
\usepackage{epstopdf}	%JWL
\newcommand{\ds}{\displaystyle}

\newcommand{\derivp} [2] {\frac {\partial #1 } {\partial #2} }
\newcommand{\deriv} [2] {\frac {\textrm{d} #1 } {\textrm{d} #2} }

\newcommand{\eq}[1] {Eq.\,(\ref{#1})}
\newcommand{\eqn} [1] {
\begin{equation} #1
\end{equation}}
\newcommand{\eqna} [1] {
\begin{eqnarray} #1
\end{eqnarray}}

\usepackage{txfonts}
\usepackage[applemac]{inputenc}
\begin{document}
\title{Mode excitation by turbulent convection in rotating stars.\\
I. Effect of uniform rotation }

\author{K. Belkacem\inst{1,2} 
\and 
S. Mathis\inst{3}
\and  
M. J. Goupil\inst{2} 
\and 
R. Samadi\inst{2}}

\offprints{K. Belkacem}

\institute
{ Institut d'Astrophysique et Géophysique, Université de Liège, Allée du 6 Août 17-B 4000 Liège, Belgium 
\and
Observatoire de Paris, LESIA, CNRS UMR 8109, F-92195 Meudon, France
\and
CEA/DSM/IRFU/Service d'Astrophysique, CE Saclay, F-91191 Gif-sur-Yvette, France
}

   \mail{Kevin.Belkacem@obspm.fr}
   \date{\today}

\abstract{We focus on the influence of the Coriolis acceleration 
on the  stochastic excitation of oscillation modes in convective regions of rotating stars. 
Our aim is to estimate the asymmetry between excitation rates of prograde and retrograde modes. 
We extend the formalism 
derived for obtaining stellar $p$- and $g$-mode amplitudes (Samadi \& Goupil 2001, Belkacem et al. 2008) 
 to include the effect of the Coriolis acceleration.   
We then study the special case of uniform rotation 
for slowly rotating stars by performing a perturbative analysis.
 This allows us to consider the cases of   the Sun and the CoRoT target HD 49933. 
We find that, in the subsonic regime, 
the influence of rotation as a direct contribution to mode driving   
is negligible compared to  the Reynolds stress contribution. 
In slow rotators, the indirect effect of the modification of
the  eigenfunctions on mode excitation is investigated   by performing 
a perturbative analysis of the excitation rates. 
The excitation of solar $p$ modes is found to be affected by rotation 
with excitation-rate asymmetries between prograde and retrograde modes of the order of several percent. 
Solar low-order $g$ modes are also affected by uniform rotation and
 their excitation-rate asymmetries are found to reach $10$ \%. 
 The CoRoT target HD 49933 is  rotating more rapidly  than the Sun ($\Omega / \Omega_\odot \approx 8$), 
  and we show that the resulting excitation-rate asymmetry is about $10$ \% for the excitation rates of $p$ modes. 
We demonstrate that $p$ and $g$ mode excitation rates   
are modified by uniform rotation through the Coriolis acceleration. 
A study of the effect of differential rotation will be presented in a forthcoming paper.}


\authorrunning{K. Belkacem et al.}

\titlerunning{Mode excitation by turbulent convection in rotating stars. I. }

\maketitle

\section{Introduction}

Internal dynamical processes in stars 
and wave excitation, propagation, and induced transport can 
be strongly influenced by rotation. Those mechanisms modify 
stellar internal structure and evolution with significant consequences for example 
for galactic evolution \citep{Maeder09}. 
The impact of rotation on stars is now studied by including  models of internal 
transport processes  in stellar evolutionary codes
\citep[see for instance][and references therein]{Talon97,Maeder00,Espinosa07,Decressin09,Maeder09}. 
Asteroseismology is also being increasingly developed with results from the  
CoRoT  \citep{Michel08,Michel08b,Appourchaux08} and KEPLER \citep[][]{JCD08} missions, 
which place constraints on stellar modeling.  
Those spatial missions allow us to study stars that are slow as well as very rapid rotators. 

Since the pioneering works of \cite{Ulrich70} and \cite{Leibacher71}, which 
led to the identification of the solar five-minute oscillations as global
 acoustic standing waves 
($p$~modes), the Sun internal structure has been determined from the
knowledge of its oscillation frequencies. 
One of the remaining key issues is the detection and identification of gravity modes 
\citep{Appourchaux00,Gabriel02,TC04,Garcia07, Mathur07,Garcia08b} for determining the rotation 
profile in the nuclear region 
\citep{Mathur08,Garcia08a}. 
Oscillation modes are indeed crucial for probing the interior of rotating stars.

Stochastic excitation of radial modes by turbulent convection has been
investigated by means of  several approaches
\citep{GK77,GK94,B92c,Samadi00I,Chaplin05}. These methods differ from each other in the
nature of the assumed  excitation sources, the adopted simplifications
and approximations, and by the way that the turbulent convection is
described \citep[see reviews by][]{Stein04,Houdek06}.  Two major
mechanisms have nevertheless been identified as driving the resonant 
$p$~modes of the stellar cavity: the first is related to the Reynolds stress tensor
and, as such, represents a mechanical source of excitation; the second
is caused by the advection of turbulent fluctuations of entropy by
turbulent motions, and as such represents a thermal source of excitation 
\citep{GK77,Samadi00I}. Samadi \& Goupil (2001) proposed 
a generalized formalism, taking the Reynolds and entropy fluctuation source terms into
account. 
The satisfying agreement between modeling and observational data for the Sun 
\citep{Belkacem06a,Belkacem06b} permitted us to go a step further and investigate the excitation 
of non-radial modes in the non-rotating case \citep{Belkacem08a,Belkacem09} and now the effect of rotation. 

Our motivation is to investigate the effect of rotation on the mode excitation rates rather than the frequencies. 
We then focus on the excitation rates of stochastically excited modes for which several issues can be addressed. For example,  
is the excitation rate of a non-axisymmetric mode ($m\neq0$) the same as for an axisymmetric one ($m=0$)? 
Are prograde and retrograde modes excited in the same manner and what are the consequences?
We pay attention to the Coriolis acceleration effects in stars, neglecting the centrifugal acceleration-induced effects such as star deformation.
Our first objective is to determine whether or not uniform rotation can drive the mode efficiently, and our 
second is to evaluate the excitation-rate asymmetry between prograde and retrograde modes induced by the perturbation of the eigenfunctions by uniform rotation. 
The effect of differential rotation on the mode excitation rates will be addressed in a forthcoming paper. 

The paper is organized as follows. Section~2 introduces the general
formalism, and a detailed derivation of the Reynolds, entropy, and rotation-induced source
terms is provided. 
In Sect.~3, the formalism is applied to solar spheroidal modes. 
The special case of slow rotators, the Sun, and the CoRoT target HD 49933 are 
then investigated, and the results are discussed.
Some conclusions are presented in Sect.~4.

\section{Turbulent stochastic excitation}

\subsection{The inhomogeneous wave equation}

We derive the inhomogeneous wave equation by taking into account  
the Coriolis acceleration and differential rotation. 
The fluid velocity field ($\vec v$) is divided into the terms
\begin{equation}
\vec v = \vec u + r\, \sin \theta \, \Omega(r,\theta) \, \vec e_\phi \, ,
\end{equation}
where $\Omega(r,\theta)$ is the rotational angular frequency assuming an axisymmetric 
 rotation, $r\, \sin \theta \, \Omega(r,\theta) \, \vec e_\phi$ is the 
velocity field associated with rotation, $\vec u$ is the velocity field associated with the turbulent 
convective motion and waves, and $\left(r,\theta,\phi\right)$ are 
the usual spherical coordinate with their associated unit vector 
basis $\left\{{{\bf e}}_{k}\right\}_{k=\left\{r,\theta,\phi\right\}}$.
The rotation axis is chosen so as
to coincide with the  $\theta = 0$ axis of the spherical coordinates system
 of an inertial frame of reference. In this work, meridional circulation is ignored.

The equation of mass conservation and motion  
in the presence of axisymmetric rotation, can be written as follows \citep[e.g., ][]{Unno89}:
\begin{align}
& \derivp{\rho}{t} + \vec \nabla \cdot (\rho \vec u) = 0 \\
& \derivp{\left(\rho \vec u\right)}{t} + \vec \nabla : \left( \rho \vec u \vec u \right) 
+ \rho \left[ \Omega \derivp{\vec u}{\phi} + 2 \, \vec \Omega \times \vec u 
+ r \sin \theta \, \vec u \cdot \, \vec \nabla \Omega \, \vec e_\phi \right] \nonumber \\
& = \rho \vec g - \vec \nabla P\,, 
\end{align}
where $\vec u$ is the velocity, $\rho$ is the density, $\vec \Omega = \Omega(r,\theta) \, \vec e_z$ 
is the rotation velocity, $\vec e_z$ is the unit vector along the rotation axis, $\vec g$ is the gravitational field, and $P$ is the pressure. We note that the centrifugal force is neglected. 

To go further, all physical quantities are divided into an equilibrium one and a 
perturbation. The subscripts 1 and 0 denote Eulerian perturbations and
equilibrium quantities, respectively,  except for velocity where the subscript
1 has been dropped for ease of notation. In the following, the
velocity field $\vec u$ is divided into two contributions, namely the
oscillation velocity ($\vec v_{\rm osc}$) and the turbulent velocity field
($\vec u_t$), such
that $\vec u= \vec v_{\rm osc} + \vec u_t$. 
Then, taking the temporal derivative of the equation of motion and using the 
mass conservation equation, one then obtains
\begin{eqnarray}
\label{wave_equation}
\left(\derivp{^2}{t^2} - \vec L_\Omega \right) \vec v_{\rm osc} + {\vec  {\cal C}_{\rm osc}} = \vec {\cal S}_t\,,
\end{eqnarray}
where $\vec L_\Omega$ is the linear operator that in presence of rotation becomes
\begin{eqnarray}
&&\vec L_\Omega = \vec \nabla \left[ \alpha_s \vec v_{\rm osc} \, \cdot \, \vec \nabla s_0 + c_s^2 \vec \nabla \left( \rho_0 \vec v_{\rm osc} \right) \right] 
- \vec g  \nabla \cdot \left( \rho_0 \vec v_{\rm osc} \right) \nonumber \\
& &- \rho_0 \Omega \frac{\partial^2 \vec v_{\rm osc}}{\partial t \partial \phi} 
- 2 \, \rho_0 \vec \Omega \times \derivp{\vec v_{\rm osc}}{t} - \rho_0 r \sin \theta \derivp{\vec v_{\rm osc}}{t} \, \cdot \, \vec \nabla \Omega \; \vec e_\phi\, .
\end{eqnarray} 
The operator $\mathcal{C}_{\rm osc}$ involves both turbulent and pulsational velocities and contributes to the 
linear dynamical damping \citep[see][for details]{Samadi00I}: 
\begin{eqnarray}
\mathcal{C}_{\rm osc} &=& \derivp{}{t} \Big[ \derivp{\left(\rho_t \vec v_{\rm osc}\right)}{t} + 2\, \vec \nabla : \left( \rho_0 \vec v_{\rm osc} \vec u_t \right) + \rho_t \Omega  \derivp{\vec v_{\rm osc}}{\phi} 
 \nonumber \\
 &+& 2 \, \rho_t \vec \Omega \times \vec v_{\rm osc} + \rho_t r \sin \theta \, \left( \vec v_{\rm osc} \cdot \vec \nabla \Omega \right) \vec e_\phi \nonumber \\
 &+& \vec \nabla \left(\alpha_s \vec v_{\rm osc} \cdot \vec \nabla s_1 
 + c_s^2 \vec \nabla \cdot (\rho_t \vec v_{\rm osc}) \right) \Big]\,.
\end{eqnarray}
Finally, the $\mathcal{S}_t$ operator that contains the source terms of the inhomogeneous wave equation (\eq{wave_equation}) is given by
\begin{align}
\label{Source}
\mathcal{S}_t &= - \derivp{}{t} \vec \nabla : (\rho_0 \vec u_t \vec u_t) + \vec \nabla (\alpha_s \vec u_t \cdot \vec \nabla s_t) + 
 \mathcal{S}_\Omega +  \mathcal{S}_M
\end{align}
and
\begin{align}
\label{termes_sources}
\mathcal{S}_\Omega &=  - \derivp{}{t} \left[ \rho_t \left( \Omega \derivp{}{\phi} \vec u_t  
 - 2 \vec \Omega \times \vec u_t %\nonumber \\
 -  r \sin \theta\, \vec u_t \cdot \vec \nabla \Omega\, \vec e_\phi  \right) \right] \\
 \mathcal{S}_M &= \derivp{}{t} (\rho_t \vec g_1) + \vec \nabla \left[c_s^2 \vec \nabla \cdot \left(\rho_t \vec u_t\right)\right] 
 - \vec g \vec \nabla \cdot (\rho_t \vec u_t) \nonumber \\
 &- \derivp{^2}{t^2} (\rho_t \vec u_t) + \mathcal{L}_t
\end{align}
where $\vec g_1$ is the perturbation of the gravitational acceleration. 
The first two terms of \eq{Source} correspond to the Reynolds stress and entropy contributions, respectively. 
The three following terms are contributions associated with rotation. 
Eventually, as shown by \cite{Samadi00I}, the terms in $\mathcal{S}_M$ do not contribute significantly to the excitation and are thus neglected as well as the linear terms\footnote{Linear terms are defined as the product of an equilibrium quantity and a fluctuating one.} ($\mathcal{L}_t$). 

\subsection{Mean square amplitude for uniform rotation}

Using \eq{wave_equation}, the next step is to determine the mean square amplitude of $\vec v_{\rm osc}$. The procedure is the same as described in \cite{Belkacem08a}. 
The wave velocity field is related to the displacement by means of the relation \citep{Unno89}
\begin{equation}
\label{v_diff}
\vec v_{\rm osc} = A\, \left[i \sigma\vec \xi - (\vec \xi \cdot \vec \nabla \Omega) r \sin \theta \, \vec e_{\phi}\right] e^{i \sigma t}\, ,
\end{equation} 
where $\sigma=\omega_0 + m \Omega$, and $\omega_0$ is the mode frequency without rotation. 
For uniform rotation, it reduces to
\begin{equation}
\label{v_nodiff}
\vec v_{\rm osc} = A\,i \sigma \vec \xi e^{i \sigma t} \, ,
\end{equation} 
where $\sigma$ is the eigenfrequency, $\vec \xi (\vec r)$ is the displacement  eigenfunction
 in absence of turbulence, and $A(t)$
is the amplitude due to the turbulent forcing. 
In the presence of rotation, the wave displacement ($\vec \xi$) is expressed as
\begin{eqnarray}
\label{decomp_harmonics}
\vec \xi (\vec r) &=& \sum_{\ell,m} \Big[ \xi_{r}^{\ell,m} \vec e_r Y_{\ell,m} + \left(\xi_{H}^{\ell,m}  \derivp{Y_{\ell,m} }{\theta} + \xi_{T}^{\ell,m} \frac{1}{\sin \theta} \derivp{Y_{\ell,m} }{\phi} \right) \vec e_H  \nonumber \\
&+& 
\left( \xi_{H}^{\ell,m} \frac{1}{\sin \theta} \derivp{Y_{\ell,m} }{\phi} + \xi_{T}^{\ell,m} \derivp{Y_{\ell,m} }{\theta} \right) \vec e_T \Big] \, ,
\end{eqnarray}
where 
 $\xi_r,\xi_H$, and 
$\xi_T$ are the 
radial, horizontal, and toroidal components of the displacement eigenfunction, respectively. 
Note that in the following we do not use the upper-scripts $^{\ell,m}$ 
on the eigenfunction components for ease of notation. Each mode is also labelled with a radial
order $n$, which we also omit. 

The power (\,$P$\,) injected into each mode with  given ($n,\ell,m$)  is  then related to  the mean-squared
 amplitude (\,$<|A|^2>$\,)  by
\begin{equation}
\label{puissance_trapping}
P = \eta <|A|^2> I ~\sigma^2 \; ,
\end{equation}
where the operator $<>$ denotes a statistical average performed
on an infinite number of
independent realizations, $\eta$ is the damping rate, and $I$ is the mode inertia.

Following \cite{Samadi00I} and \cite{Belkacem08a}, 
one then obtains the mean square amplitude for each mode as 
\begin{equation}
\label{mean_square_amplitude_trapping}
<|A|^2> = \frac{1}{8 \eta \left(\sigma I\right)^2} \left( C_R^2 + C_S^2 + C_\Omega^2 + C_{c}  \right) \;,
\end{equation}
where $C_R^2$ is the Reynolds stress contribution, $C_\Omega^2$ contains the contributions related to the 
Coriolis acceleration, the Doppler term, and one related to the differential rotation, $C_S^2$ corresponds to 
entropy fluctuation contributions, while $C_{c}$ represents the cross-source terms, i.e.,  the 
interferences between the different source terms. 

\subsection{Reynolds stress contribution}
\label{terme_source_reynolds}

Following the formalism of \cite{Belkacem08a}, we develop the turbulent 
Reynolds contribution (see Appendix~\ref{App:Reynolds} for a detailed derivation), 
which becomes for a given ($\ell,m$)
\begin{eqnarray}
\label{C2R_ref}
C_R^2 & =  &  16\pi^{4}   \int  \textrm{d}r   \, r^2 \rho_0 \; R(r)~ S_R(\sigma)\; ,
 \end{eqnarray}
and
\begin{eqnarray}
\label{gammabeta}
R (r) &=&   {16\over 15} ~  \left| \deriv{\xi_r}{r}  \right|^2  +  {44\over 15} ~    \left| \frac{\xi_r}{r}  \right|^2
+   \frac{4}{5} \left( \frac{\xi^*_r}{r} \deriv{\xi_r}{r} + {\rm c.c.} \right) \nonumber \\ 
&+&  ~ L^2 \left[ {11\over 15} ~ \left( \left| \mathcal{A}  \right|^2 + \left| \mathcal{B}  \right|^2 \right) - {22 \over 15} \left(\frac{\xi_r^* \xi_H}{r^2} +{\rm c.c.}\right) \right] \nonumber \\
&+&  \left| \frac{\xi_H}{r}  \right|^2 
\left( \frac{16}{15} L^4+ \frac{8}{5} {\cal F}_{\ell,\vert m \vert} - \frac{2}{3} L^2 \right) \nonumber \\
&+&  \left| \frac{\xi_T}{r}  \right|^2  \left( \frac{11}{5} L^2 (L^2-2) - \frac{8}{5} {\cal F}_{\ell,\vert m \vert} - \frac{2}{3} L^2 \right) \nonumber \\
 &-&{2\over 5} L^2 \left(\deriv{\xi^*_r}{r} {\xi_H \over r} 
+ {\rm c.c.}\right)\,,
\end{eqnarray}
where  
\begin{equation}
\mathcal{A} = \deriv{\xi_H}{r} +\frac{1}{r}(\xi_r-\xi_H)\quad\hbox{and}\quad\mathcal{B} =  \deriv{\xi_T}{r} - \frac{\xi_T}{r}\,,
\end{equation}
and
\begin{equation}
{\cal F}_{\ell,|m|} =  \frac{|m| (2 \ell + 1)}{2} \left[ \ell(\ell+1) - (m^2 + 1 )\right]
\end{equation}
and $L^2=l\left(l+1\right)$. Furthermore,
\begin{equation}
S_R(\sigma) = \,\int  \frac {\textrm{d}k} {k^2 }~E^2(k) ~\int \textrm{d}\omega 
~\chi_k( \omega + \sigma) ~\chi_k( \omega )\, ,
\label{fct_source}
 \end{equation}
where $(\vec k,\omega)$ 
are the wave number and frequency of the
turbulent eddies, and $E(\vec k,\omega)$ is the turbulent kinetic energy spectrum, which is expressed as the product $E(\vec k) \,\chi_k(\omega)$  for isotropic turbulence \citep{Stein67}. A detailed
discussion of the temporal correlation function ($\chi_k$) is addressed in \citet{Samadi02I}.

Note that in absence of rotation (\emph{i.e.}, $\Omega=0$), 
the toroidal component of the eigenfunction $\xi_T$ vanishes 
in \eq{decomp_harmonics} and for $C_R^2$ and $R(r)$ we recover the expressions given by Eqs.~(22) and (23) of \cite{Belkacem08a}. 
From \eq{gammabeta}, additional terms are found to appear through the toroidal component of the eigenfunction. All are found to be positive regardless of $\ell$ and $m$, implying an increase in the excitation rates. 

We emphasize that rotation is understood to create anisotropies in the Reynolds stress tensor, then off-diagonal terms \citep[e.g., ][]{Kumar95,Miesch05}. 
An adapted spectral description of turbulent convection
 including the effect of rotation is thus required to compute \eq{fct_source} and  is beyong the 
 scope or our study. 

\subsection{Entropy fluctuation contribution}

As shown by \cite{Samadi00I} and \cite{Belkacem06b}, the Reynolds stress contribution 
is not the unique source of excitation but one has to account for the excitation by 
the entropy contribution to reproduce the excitation rates for solar radial $p$~modes. 

Following \cite{Belkacem08a}, the entropy source term 
depends on the mode compressibility that can be estimated as
\begin{equation}
\label{C_S_ref3}
\int_{\bar \Omega} \, \textrm{d}\bar\Omega \, Y_\ell^m \, \vec \nabla \cdot \vec \xi = 
 \frac{1}{r^2} \, \deriv{}{r} \left (  r^2  \xi_r \right ) - \frac{L^2 } {r} \, \xi_H \; . \;
\end{equation}
where $\bar \Omega$ is the solid angle, and the spherical harmonics are normalized  
following \eq{normalisation}.

Hence, from Eqs.~\ref{decomp_harmonics} and \ref{C_S_ref3} the divergence of the toroidal component, which is the curl of the spherical harmonic, vanishes. 
Consequently, one obtains the same result as for taking only the poloidal contribution into account. 
The final expression for the contribution of entropy fluctuations remains the same as in  
\cite{Belkacem08a}, \emph{i.e.},
\eqn{
\label{C_S_ref}
C_S^2  = \frac{4 \pi^3 \, \mathcal{H}}{\sigma^2}  \, 
\int \textrm{d}^3 x_0 \, \alpha_s^2 \, \left ( A_\ell + B_\ell \right ) \,
\mathcal{S}_S(\sigma) \, ,
}
where $\mathcal{H}$ is the anisotropy factor introduced in \cite{Samadi00I}, which,
 for the current
assumption (isotropic turbulence),  is equal to  $ 4/3 $.  In addition,
\eqna{
\label{C_S_ref2}
A_\ell & \equiv & \frac{1 }{r^2}  \,  \left|  D_\ell \, 
\deriv{\left( \ln \mid \alpha_s \mid \right)}{\ln r}
 - \deriv{D_\ell }{\ln r}  \right|^2\,,\label{eqn:Al}
\\
\label{Bell}
B_\ell & \equiv &  \frac{1 }{r^2} \, L^2 \, \left| D_\ell \right| ^2\,, \\
D_\ell & = & \frac{1}{r^2} \, \deriv{}{r} \left (  r^2  \xi_r \right ) - \frac{L^2 } {r} \, \xi_H
\label{eqn:Bl}}
where
\begin{equation}
\mathcal{S}_S(\sigma)  \equiv  \int \frac{\textrm{d}k}{k^4}\,E(k)
\, E_s(k) \, \int \textrm{d}\omega \,
\chi_k(\sigma+\omega)\,  \chi_k(\omega)\,.\label{eqn:FS}
\end{equation} 
 In contrast to the Reynolds contribution expression \eq{gammabeta}, the entropy one is not directly modified by rotation. Nevertheless, this contribution can be influenced indirectly by means of the modification of the radial and horizontal components of the eigenfunctions ($\xi_r$ and $\xi_H$) by the Coriolis acceleration.

\subsection{Rotational contributions}
\label{terme_source_rot}

The rotational contributions in the inhomogeneous equation (Eqs.~\ref{wave_equation} and \ref{termes_sources}) are
\begin{itemize}
\item the contribution related to the Coriolis acceleration
\begin{align}
\label{coriolis}
- \derivp{}{t} \left( 2 \, \rho_t \vec \Omega \times  \vec u_t \right) 
= - 2 \vec \Omega \times \derivp{}{t} \left( \rho_t \vec u_t \right) \, , 
\end{align}
where we neglect the time variations in the angular velocity on a dynamical time scale. 
\item the contribution related to the Doppler shift
\begin{align}
\label{doppler}
\derivp{}{t} \left( \rho_t \Omega \derivp{\vec u_t }{\phi} \right) 
= \Omega \derivp{}{t} \left( \rho_t \derivp{\vec u_t }{\phi} \right) \, ,
\end{align}

\item the contribution related to the differential rotation
\begin{align}
\label{differential}
-\derivp{}{t} \left( \rho_t r \, \sin \theta \, \vec u_t \cdot \vec \nabla \Omega \right)  \vec e_\phi= 
r \sin \theta \, \derivp{\rho_t \vec u_t}{t} \cdot \vec \nabla \Omega\, \vec e_\phi .
\end{align}
\end{itemize}
In this paper, we consider only uniform rotation, hence the last contribution (\eq{differential}) vanishes. Nevertheless, all contributions, \emph{i.e.}, from \eq{coriolis} to \eq{differential}, are proportional to the perturbed mass flux $\rho_t \vec u_t$.    
A dimensional analysis \citep[see][for details]{Samadi00I} shows that  
all those terms then scale as the Mach number to the third ($\mathcal{M}^3$). 
Compared to the Reynolds contribution, which scales as $\mathcal{M}^2$, all rotational contributions are negligible in the subsonic regime. For the Sun, this conclusion remains valid even for the uppermost layers where  $\mathcal{M} \approx 0.3$. 
In addition, the rotational velocity appears from \eq{coriolis} to \eq{differential} introducing the ratio $\Omega / \sigma$, which is very small for slow rotators.  

Eventually, one obtains (see \eq{C2R_ref}, \eq{C_S_ref}, and \eq{coriolis} to \eq{doppler})
\begin{align}
C_R^2 &= \mathcal{O}(\mathcal{M}^4)  
\gg C_S^2 = \mathcal{O}(\mathcal{M}^6) 
\gg C_{R\Omega} = \mathcal{O}(\mathcal{M}^5)  \left(\frac{\Omega}{\sigma}\right)  \nonumber \\
&\gg C_\Omega^2 = \mathcal{O}\left(\mathcal{M}^6 \right) \; \mathcal{O}  \left(\frac{\Omega}{\sigma}\right)^2 \, ,
\end{align}
where $C_{R\Omega}$ is the coupled source term associated with the Reynolds stress and rotational contributions.  
Consequently, in the following only the Reynolds stress contribution will be considered. 

\subsection{Final balance}
\label{balance}

We have shown in Sects.~\ref{terme_source_reynolds} to \ref{terme_source_rot} 
that in the presence of uniform rotation, 
the Reynolds term contribution (\eq{C2R_ref}) remains the most dominant in the subsonic regime. It can be influenced by uniform rotation in three ways; 
\begin{itemize}
\item The turbulent velocity field can be modified by the Coriolis acceleration, hence affecting the Reynolds contribution in \eq{C2R_ref} by the source term (\eq{fct_source}). 
\item The toroidal component of the eigenfunction introduces additional terms in \eq{gammabeta}. 
\item Eventually, the poloidal components of the eigenfunctions are modified by the Coriolis acceleration and  will influence the Reynolds contribution in terms of \eq{C2R_ref}. 
\end{itemize}

\section{Application to spheroidal modes of slow rotators}
\label{slow_rotator}

As mentioned in Sect.~\ref{balance}, the velocity field can be modified by the Coriolis acceleration. 
However, for slow rotators the rotation rate does not significantly affect  
the turbulent field in the upper convective region where modes are excited, 
provided that the ratio of the convective frequency to the rotation rate is higher than unity. 
For the Sun, this requirement is fulfilled in the entire  
convective region except in the deepest layers, near the interface with the radiative region. 
Nevertheless, the contribution of these deep layers do not contribute significantly 
to the excitation rates for the modes considered here, \emph{i.e.}, low-order $g$ modes 
and $p$ modes. 
Hence, in the following we assume that the turbulent field, and its spectral dependence, 
are not affected by uniform rotation. Note, however, that for lower frequencies, 
and especially asymptotic gravity-modes, this approximation is no 
longer valid since a significant contribution to the mode excitation 
comes from the deeper convective layers \citep{Belkacem09}. 

We then consider the effect of the perturbation of the 
mode excitation rates by the Coriolis acceleration associated with the modification of the eigenfunctions. 
In this framework, we use a perturbative approach, 
which is valid for slow rotators and particularly for the Sun 
since we restrict our investigation to rather high-frequency $p$ and $g$ modes. 
The ratio of the mode frequency to the rotation rate is still higher than unity. In the Sun, for a typical $p$ mode 
at $\nu=3 \,$mHz, one has $\omega_0/2 \Omega \approx 3 \times 10^3 \gg 1$ in 
the convective region where modes are excited, 
and for a solar $g$ mode at $\nu=100\,\mu$Hz, this ratio remains high at $\omega_0/2 \Omega \approx 100  \gg 1$. This allows us to use a perturbative approach. 

\begin{figure}[t]
\begin{center}
\includegraphics[height=6cm,width=9cm]{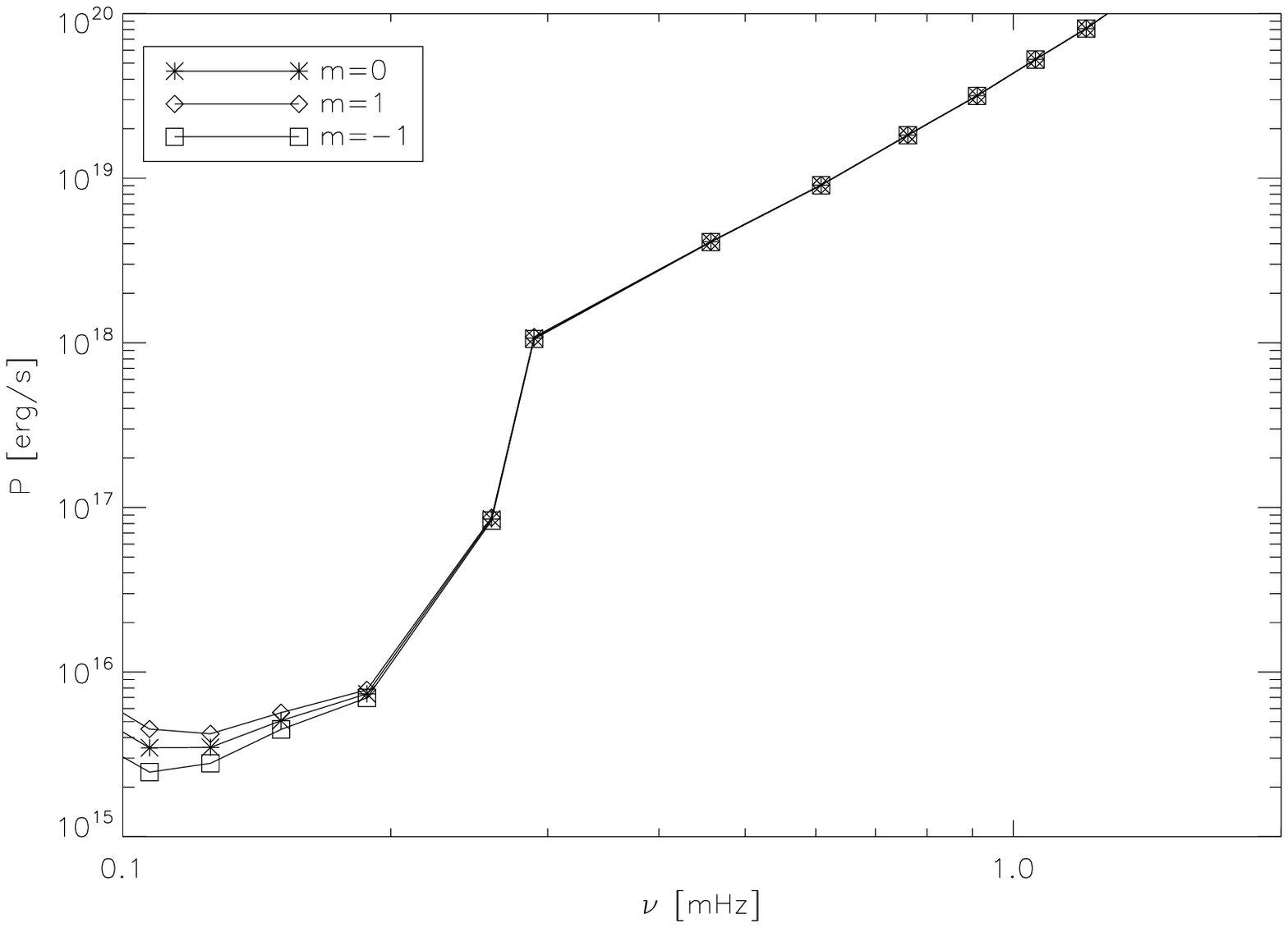}\\
\includegraphics[height=6cm,width=9cm]{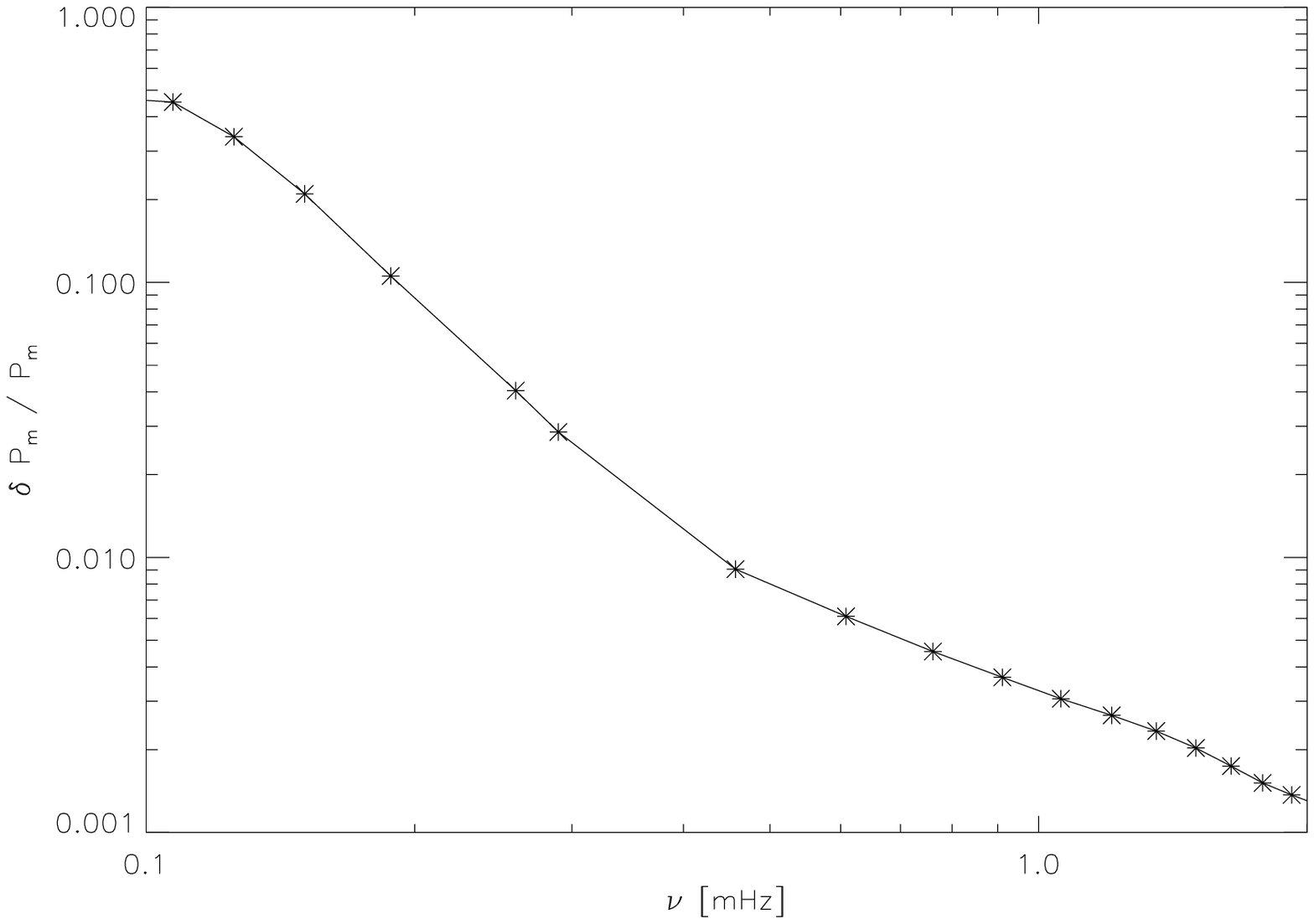}
\caption{{\bf Top:} Excitation rates for the mode $\ell=1$ and $m=\{-1,0,1\}$, computed using Eqs.~(\ref{puissance_trapping}), 
(\ref{mean_square_amplitude_trapping}), and (\ref{C2R_ref}) and using the same 
solar model as detailed in \cite{Belkacem08a}. The depression in the excitation rates 
at $\nu \approx 300 \, \mu$Hz is associated with the mixed nature of the modes that consequently produce  
a higher inertia thus  a lower $P$. 
{\bf Bottom:} Bias between prograde and retrograde modes, defined 
in \eq{deltaP}, 
for the same modes as for the figure in the top panel.  
}
\label{figure_P_ell0}
\end{center}
\end{figure}

\subsection{Perturbation of the mode excitation rates}

Our aim is to derive an analytical expression of excitation rates influenced by uniform rotation 
using a perturbative method. 
Following the classical method given in \cite{Unno89}, we develop the eigenfunction
for a given $n,\ell,m$ to first order
\begin{align}
\xi_{r}&=\xi_{r;n,\ell}^{\left(0\right)}+m\left(\frac{2\Omega}{\omega_0}\right)\sum_{n^{'} \neq n}C_{n^{'},n,\ell}\; \xi_{r;n^{'},\ell}^{\left(0\right)}\, , \\
\xi_{H}&=\xi_{H;n,\ell}^{\left(0\right)}+m\left(\frac{2\Omega}{\omega_0}\right)\sum_{n^{'} \neq n} C_{n^{'},n,\ell} \; \xi_{H;n^{'},\ell}^{\left(0\right)} \, ,
\label{xith}
\end{align}
and for the toroidal part
\begin{align}
\xi_{T}&= i \left(\frac{2\Omega}{\omega_0}\right) \Big\{ \ell D_{\ell,m}\left[\xi_{r;n,\ell-1}^{\left(0\right)}-\left(\ell-1\right)\xi_{H;n,\ell-1}^{\left(0\right)}\right]\nonumber\\
& -\left(\ell+1\right)D_{\ell+1,m}\left[\xi_{r;n,\ell+1}^{\left(0\right)}+\left(\ell+2\right)\xi_{H;n,\ell+1}^{\left(0\right)}\right] \Big\} \, ,
\label{xiazc}
\end{align}
where 
\begin{equation}
D_{\ell,m}=\frac{1}{\ell^2}\sqrt{\frac{\ell^2-m^2}{4\ell^2-1}}.
\end{equation}
The expression for $C_{n^{'},n,\ell}$ is given in Appendix~\ref{first_order}, and $\xi_{r;n,\ell}^{\left(0\right)}, \xi_{H;n,\ell}^{\left(0\right)}$  are the radial and horizontal components of the eigenfunction in absence of rotation. 

As pointed out by \cite{DG92}, the convergence properties of the sum involved in \eq{xirc} and \eq{xithc} are unclear. 
 As shown by \eq{Cnl}, this is particularly problematic for a dense spectrum such as high-order gravity modes in the Sun. 
 We nevertheless use it for convenience. An alternative exists \citep{DG92}, which consists of computing a modified eigenvalue problem. However, this second possibility makes it more  difficult to identify contributions to the excitation rates.

Inserting the decomposition (Eqs.~\ref{xirc} to \ref{xiazc}) into Eqs.~\ref{C2R_ref} and \ref{gammabeta}, we obtain (see Appendix~\ref{first_order} for the detailed calculation)
\begin{equation}
\label{perturb_CR2}
P_m =P^{(0)} + m \, \left(\frac{2\Omega}{\omega_0}\right) P_{|m|}^{(1)} \, ,
\end{equation}
where
\begin{equation}
\label{perturb_reynolds}
P^{(0)}=\frac{4\pi^3}{8 I} \int{\rm d}m\,  R^{\left(0\right)} \, S_{R}\left(\omega_{0}\right)\,,
\end{equation}
\begin{equation}
\label{perturb}
P_{|m|}^{(1)}=\frac{4\pi^3}{8 I} \int{\rm d}m \, R_{|m|}^{\left(1\right)} \, S_{R}\left(\omega_{0}\right)\,,
\end{equation}
and $R^{\left(0\right)}$ and $R^{\left(1\right)}$ corresponds to the perturbative expansion of 
$R$ (\eq{gammabeta}) given in Appendix~\ref{first_order}. 
Note that the zeroth-order terms ($^{0}$) correspond 
to the case without rotation. Only the first order in $(2\Omega / \sigma)$ 
is considered. Accordingly, the contributions of $\xi_T$ in \eq{C2R_ref} are neglected because  
they are of second order. 

We now define the excitation rates asymmetry, between prograde and retrograde modes to first order such as
\begin{align}
\label{deltaP}
\frac{\delta P_{m}}{P_{m}} = \frac{P_{|m|} - P_{-|m|}}{P_{|m|}} \approx  2\,  m \left(\frac{2 \Omega}{\omega_0}\right) \,\left( \frac{P_{|m|}^{(1)}}{P^{(0)}}\right) \, .
\end{align}
From \eq{deltaP}, two factors contribute to the asymmetry namely the ratio 
$(2\Omega / \omega_0)$ and $P_{|m|}^{(1)} / P^{(0)}$. They are discussed in the following sections. 

\subsection{Application to slow rotators}

\begin{figure}[t]
\begin{center}
\includegraphics[height=6cm,width=9cm]{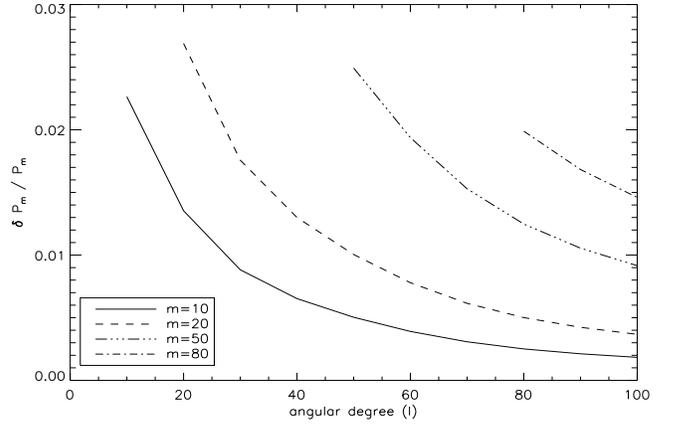}
\caption{Bias between prograde and retrograde modes, defined 
in \eq{deltaP}, 
for four values of the azimuthal order $m$ and the radial order $n=5$ as a function of the angular degree 
$\ell$. The computations are performed in the same manner as for Fig.~\ref{figure_P_ell0}.}
\label{figure_P_ell1}
\end{center}
\end{figure}

\subsubsection{The Sun}

Using the same numerical computation as described in \cite{Belkacem08a}, we apply this 
formalism (\eq{perturb_CR2} - \ref{perturb}) to the solar case. 
Figure~\ref{figure_P_ell0} displays the mode excitation rates for the $\ell=1$, $p$ and $g$ modes. 
It also presents the ratio $\delta P_{m} / P_{m}$, defined in \eq{deltaP}, which emphasizes 
the effect of the mode excitation rate asymmetry between the prograde and retrograde modes. 

It turns out that the excitation rates of acoustic modes are modified 
with an excitation rate asymmetry of the order of the percent,  which 
increases toward $g$ modes.   
We find that the variation in the mode excitation-rate asymmetry with frequency 
is caused by the term $2\Omega / \omega_{0}$ 
in \eq{deltaP}, while the ratio $P_{|m|}^{(1)} / P^{(0)}$
remains of the order of the value of one.
$P_{|m|}^{(1)}$ is dominated by the first term in \eq{perturb_cr}, 
which corresponds to the contribution of the radial component 
of the eigenfunction, for $p$ modes. 
For $g$ modes, the horizontal component of the eigenfunction 
is also of importance and contributes significantly to 
$P_{|m|}^{(1)}$. 

For higher values of  the angular degree ($\ell$), as 
shown by Fig.~\ref{figure_P_ell1}, there are two effects. 
First, the higher the azimuthal order $m$, the higher the mode excitation rates  
asymmetry, at fixed $\ell$.
 This is explained by the perturbation of the mode excitation rates
  being proportional to $m$ in \eq{perturb_CR2}. 
Second, at fixed $m$, the higher the angular degree, the lower the mode excitation-rate   
asymmetry. 
This behavior comes from the frequency shift of high-$\ell$ modes, since at fixed radial order,  
the higher the angular degree the higher the mode frequency. Hence, for the same radial order
 the ratio $2 \Omega / \omega_{0}$ will 
decrease with the angular degree explaining the behavior in Fig.~\ref{figure_P_ell1}.

\subsubsection{The CoRoT target HD 49933}

We now consider more rapid rotators, such 
as HD 49933. This is an F5 V main-sequence star observed twice by the CoRoT mission\footnote{The CoRoT space mission, launched on December 27th 2006, has been developed and is operated by CNES, with the contribution of Austria, Belgium, Brazil , ESA (RSSD and Science Program), Germany and  Spain.}, first during $62$ days and more recently for more than $150$ days. The unprecedented photometric precision achieved by the CoRoT mission \citep{Michel08,Auvergne09} makes this star a good candidate for the detection of mode excitation-rate asymmetry, which requires, as previously mentioned, accurate measurements. 
This star, indeed, exhibits a surface rotation period that is shorter than that of the Sun, $P_\Omega \approx 3.4$ days (\emph{i.e.}, $\Omega / \Omega_\odot \approx 8$) as shown by \cite{Appourchaux08}, but still slow enough to ensure that the perturbative approach is valid. 

\begin{figure}[t]
\begin{center}
\includegraphics[height=6cm,width=9cm]{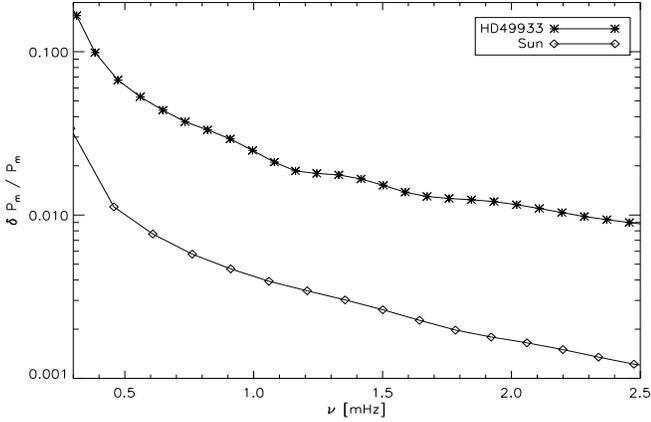}
\caption{Bias between prograde and retrograde modes, for the $\ell=1$ modes 
of the star HD49933 ($P_\Omega$=3.4 d) and the Sun ($P_\Omega$=28 d).}
\label{figure_P_ell2}
\end{center}
\end{figure}

In Fig.~\ref{figure_P_ell2}, we present the same ratio as in Fig.~\ref{figure_P_ell1} for the $\ell=1$, $p$ 
modes using a model of HD 49933 that matches
 the seismic constraints derived by \cite{Appourchaux08} \citep{Goupil09}.
  The asymmetry between the excitation rates of $m=1$ and $m=-1$ modes is
   found to reach up to  $10\%$. In terms 
   of mode excitation-rate asymmetry, the differences between the Sun and HD 49933  is due to a higher  ratio $(2\Omega / \omega_0)$ in \eq{deltaP}. 

This demonstrates that an asymmetry in terms of mode excitation rates is more likely to be observable for more rapid rotators than the Sun, even if, in contrast to the Sun, only low-$\ell$ modes are observed. 

\section{Conclusion and perspectives}

\subsection{Conclusion}

We have derived a formalism that models the stochastic 
excitation of oscillation modes by convective motions in uniformly 
rotating stellar regions.  
We have shown that the driving terms, due to rotation, that appear in the inhomogeneous wave equation are negligible with 
respect to the Reynolds stress contribution. We demonstrate that the dominant contribution 
to the excitation rates then comes from the modification of the eigenfunction by the Coriolis acceleration.

The formalism is then applied to low-order $g$ modes and $p$ modes of slow rotators, 
and in particular the Sun and the CoRoT target HD 49933. 
For the Sun, a bias between pro- and retrograde waves is found in 
the excitation rates. For $g$ modes, this bias can reach values of up to  $10$ \%. For low-$\ell$ $p$ modes, this bias is found to be of the order of a percent. 
The detection of the mode excitation-rate asymmetry of individual $p$ modes is not yet 
possible since the observational error bars obtained, for instance by GOLF, are around 
$20$ \% \citep{Belkacem06b}, while we search for a physical effect of only several percent. 

For more rapid rotators, such as HD49933, we find that the excitation-rate asymmetry 
of low-$\ell$ ($\ell=1$) $p$ modes can reach up to $10$ \%. 
However, this value is achieved at low frequency, where seismic measurements are generally dominated 
by the granulation background. In the case of the CoRoT target HD49933, detection of acoustic modes
 is limited to the frequency domain $\nu \in [1.2 ; 2.5]$ mHz. In this frequency domain, 
 the asymmetry in $P$ is no greater than 2\%. In contrast, the 1-$\sigma$ uncertainties 
 associated with $P$ are in the range 30\% -  80\%, depending on the frequency, for the observations 
 completed during the CoRoT initial run \citep{Appourchaux08,Samadi09b}. 
 For the second set of observations of HD 49933 by CoRoT, the 1-$\sigma$ uncertainties associated with $P$ is expected to be lower, \emph{i.e.}, in the range  20\% - 40 \% (Benomar, private communication). 
Furthermore, current seismic analyses \cite[e.g.,][]{Appourchaux08} do not reproduce 
 individual mode multiplets but  assume a fixed amplitude ratio of the different mode multiplets or even assume a fixed amplitude ratio of the different $\ell$ degree. 
Therefore, despite the high  precision of the CoRoT instrument, it is presently not possible
 to constraint $P$ for an individual  mode multiplet (\emph{i.e.}, for a given value of $\ell$ and $m$). 
Concerning the Kepler instrument, its performance in terms of photon noise level 
is expected to be a factor five lower in terms of power compared to that achieved for the brightest stars 
of the CoRoT mission \citep{Chaplin08}. On the other hand, Kepler will observe the seismic 
targets over a much longer period (around 4 years) than the CoRoT mission, which 
we hope will permit us to constrain individual mode multiplets. 
 Another way to proceed is to consider a sum of $P$ for a given $m$ so as to reduce the actual 
observational errorbars in both the Solar case and that of HD49933. 

We note that mode amplitude is a balance
between driving and damping. Therefore, asymmetries in mode amplitudes
cannot be inferred only from excitation rates since
some possible asymmetry in the mode damping rate can arise. 
This is not investigated here but left to future work.

\subsection{Perspectives}

 The effect of uniform rotation on the mode amplitude excitation rates presented here is exploratory work 
 that requires further investigation and theoretical developments.  

Stellar convection zones are differentially rotating. 
Therefore, the next step would be to take the differential rotation into account 
 in both the radial and the latitudinal 
directions. In contrast to uniform rotation, a consideration at \eq{v_diff} permits us to understand 
that the driving source terms 
in the inhomogeneous wave equation \eq{wave_equation} are modified by 
differential rotation. 
The effect of differential rotation on mode excitation rates is the scope 
of an upcoming paper.

The regime of rapid rotation should also be addressed. 
The formalism must be adapted to the specific geometry of those stars, since 
spherical coordinates become inappropriate and one may have to take the star deformation 
into account. 
In addition, the eigenfunctions and frequencies have to be derived from 
an adapted non-perturbative method \citep[e.g., ][]{Reese06} 
since rotation can strongly modify both the eigenfunctions and
 the stellar structure of a star \citep{Rieutord97b,Dintrans00,Reese06}. 
Furthermore, in such a regime new types  of waves appear that 
deserve a further study, such as inertial waves and  gravito-inertial
 waves \citep{Rieutord97b,Dintrans00,Mathis08}.  
In addition, in contrast to slow rotators, the turbulent field is also affected 
by rotation and the spectral description of turbulence must be taken it into account. The assumption of isotropic turbulence is should then be excluded and the spectral properties of the turbulent 
field be specified.  
Numerical simulations can be of some help. As done for solar $g$ modes using the ASH code 
\citep{Belkacem09}, it is possible to assess the turbulent properties of these rotators 
from numerical simulations.

\begin{acknowledgements}
K.B. acknowledges financial support from Li\`ege University through the subside f\'ed\'eral pour la recherche.
\end{acknowledgements}

%\bibliographystyle{aa}
%\bibliography{bib.bib}

\newpage

\appendix

\section{Detailed expressions for the Reynolds source term}
\label{App:Reynolds}

%%%%%%%%%%%%%%%%%%%%%%%%%%%%%%%%%%%%%%%%%%%%%%%%%%%%%
The eigenfunctions (\,$\vec \xi$\,) are developed in spherical coordinates
$(\vec e_r, \vec e_\theta, \vec e_\phi)$
and expanded as a sum over spherical harmonics.
Hence the fluid displacement eigenfunction for 
a mode with given $\ell,m$ is written as
\begin{eqnarray}
\label{decomp_harmonics2}
\vec \xi (\vec r) &=& \sum_{\ell,m} \Big[ \xi_{r;\ell,m} \vec e_r Y_{\ell,m} + \left(\xi_{H;\ell,m} \derivp{Y_{\ell,m} }{\theta} + \xi_{T;\ell,m} \frac{1}{\sin \theta} \derivp{Y_{\ell,m} }{\phi} \right) \vec e_H  \nonumber \\ 
&+& 
\left( \xi_{H;\ell,m} \frac{1}{\sin \theta} \derivp{Y_{\ell,m} }{\phi} + \xi_{T;\ell,m} \derivp{Y_{\ell,m} }{\theta} \right) \vec e_T \Big] \, ,
\end{eqnarray}
where the spherical harmonics ($Y_{\ell,m} (\theta,\phi)$) are normalized according to 
\begin{equation}
\label{normalisation}
\int \frac{d\bar\Omega}{ 4 \pi} \,  Y_{\ell,m} ~Y^*_{\ell,m}   = 1 
\end{equation}
with $\bar\Omega$ being the solid angle ($\textrm{d}\bar\Omega = \sin \theta \, \textrm{d} \theta \, \textrm{d} \phi$).\\

The Reynolds stress contribution  can be written as \citep[see][for details]{Belkacem08a}
\begin{eqnarray}
\label{cr_interm}
C_R^2 & = &\pi^{2}   \int  {\textrm{d}^3 x_0}  \,
\left (\rho_0^2 \,  b^*_{ij} b_{lm} \right )
\int \textrm{d}^3k \, \int \textrm{d}\omega \,  \nonumber \\
&\times& \left (  T^{ijlm} + T^{ijml} \right )
\frac {E^2(k)} {k^4 }  \, \chi_k( \sigma+\omega) \, \chi_k(\omega)
\label{eqn:C2R_2}
\end{eqnarray}
where
\begin{align}
\label{Tijlm}
T^{ijlm} &= \left( \delta^{il}- \frac {k^i k^l} {k^2}  \right)   \left( \delta_{jm}-
\frac {k^j k^m} {k^2}  \right) \\ 
\label{eqn:bij}
b_{ij} &\equiv \vec {e}_i \, \cdot \, \left ( \vec \nabla_0 \, : \, \vec \xi \right ) \, \cdot
\,  \vec {e}_j \, ,
\end{align}
where the double dot denotes the tensor product.

To compute the coefficients $b_{ij}$ in \eq{cr_interm}, we follow the procedure derived by \cite{Belkacem08a}, which infers that 
\eqn{
\begin{array}{lll}
\vspace{0.2cm} b_{rr} & = &  \ds \sum_{\ell,m} \left\{\left ( \deriv{\xi_r }{r} \right) \, Y_{\ell,m}\right\},  \\
\vspace{0.2cm} b_{r \theta}  & =  & \ds \sum_{\ell,m} \left\{\left(  \deriv{\xi_H}{r}  \right)  \,
\derivp{ Y_{\ell,m}}{\theta}   +  \left(  \deriv{\xi_T}{r}  \right)  \,
\frac{1}{\sin \theta} \derivp{ Y_{\ell,m}}{\theta}\right\}, \nonumber  \\
\vspace{0.2cm} b_{r \phi}  & =  & \ds   \sum_{\ell,m} \left\{\left(   \deriv{\xi_H}{r}
  \right) \frac{1}{\sin \theta }
\,  \derivp{Y_{\ell,m} }{\phi}    -  \left(  \deriv{\xi_T}{r}  \right)  \,
\derivp{ Y_{\ell,m}}{\theta}\right\},    \nonumber \\
%%%%%%%%%%%%%%%
\vspace{0.2cm} b_{\theta r} & = & \ds \sum_{\ell,m}\left\{ \frac{1}{r} (\xi_r -  \xi_H)
 \derivp{Y_{\ell,m}}{\theta}  - \frac{\xi_T}{r} \frac{1}{\sin \theta} \derivp{Y_\ell^m}{\phi}\right\}, \nonumber \\
\vspace{0.2cm}  b_{\theta \theta} & = & \ds \sum_{\ell,m}\left\{  \frac{\xi_H}{r}
\left ( \derivp{^2  Y_{\ell,m}}{\theta^2} \right ) +  \frac{\xi_r}{r} Y_{\ell,m}   + \frac{\xi_T}{r} 
\derivp{}{\theta}\left( \frac{1}{\sin \theta} \derivp{Y_\ell^m}{\phi}\right)\right\},  \nonumber \\
\vspace{0.2cm}  b_{\theta \phi} & = &  \ds  \sum_{\ell,m}\left\{ \frac{ \xi_H}{r} 
\derivp{}{\theta}\left [\frac{1}{\sin \theta }
\derivp{Y_{\ell,m}}{\phi} \right ] - \frac{\xi_T}{r}  \derivp{^2 Y_\ell^m}{\theta^2}\right\},
  \nonumber \\
\vspace{0.2cm}  b_{\phi \theta} & = &  \ds \sum_{\ell,m}\left\{ \frac{ \xi_H}{r} 
\derivp{}{\theta}\left [\frac{1}{\sin \theta }
\derivp{Y_{\ell,m}}{\phi} \right ] + \frac{\xi_T}{r} \left( \derivp{^2 Y_\ell^m}{\phi^2} 
+ \frac{\cos \theta}{\sin \theta} \derivp{Y_\ell^m}{\theta} \right)\right\},
  \nonumber \\
%%%%%%%%%%%%%%%
\vspace{0.2cm}  b_{\phi r} & = & \ds  \sum_{\ell,m}\left\{ \frac{1}{r} (\xi_r - \xi_H) \frac{1}{ \sin \theta }
\derivp{Y_{\ell,m}}{\phi}  + \frac{\xi_T}{r} \derivp{Y_\ell^m}{\theta}\right\}, \nonumber \\
\vspace{0.2cm}  b_{\phi \phi} & = & \ds \sum_{\ell,m}\left\{ \frac{\xi_r}{r} Y_{\ell,m}  + \frac{\xi_H}{r} 
  \ds \left [ \frac{1}{\sin^2 \theta }\left ( \derivp{^2 Y_{\ell,m}}{\phi^2}  \right )
+   \frac{\cos \theta}{\sin \theta} ~
\left ( \derivp{Y_{\ell,m}}{\theta} \right ) \right ]\right. \nonumber \\
&+&{\left. \ds \frac{\xi_T}{r} \left( \frac{\cos \theta}{\sin^2 \theta} \derivp{Y_\ell^m}{\phi} 
- \frac{1}{\sin \theta} \frac{\partial^2 Y_\ell^m}{\partial \phi \partial \theta} \right)\right\}}.
\label{eqn:bij_4}
\end{array}
}
The contribution of the Reynolds stress can thus be written as:
\begin{eqnarray}
C_R^2 & =  &  4\pi^{3}   \int  \textrm{d}m  \,\int \textrm{d}k ~\int \textrm{d}\omega \,  \; R(r, k)
 \nonumber \\  &\times& \,
\frac {E^2(k)} {k^2 }  \chi_k( \omega + \sigma) \chi_k( \omega )  \; ,
\label{eqn:C2R_3}
 \end{eqnarray}
where we have defined $\textrm{d}m= 4\pi r^2 \rho_0 \textrm{d}r$. Using the Einstein summation convention 
\begin{eqnarray}
R (r, k) = \int {\textrm{d}\bar\Omega\over 4 \pi} ~ \int  {\textrm{d}\Omega_k \over 4 \pi}\, b^*_{ij}
\, b_{lm} \, \left ( T^{ijlm} + T^{ijml} \right )\,.
\end{eqnarray}
Because $T^{ijlm}=T^{jiml}$, it is easy to show that 
\begin{eqnarray}
R (r, k) = \int {\textrm{d}\bar\Omega\over 4 \pi} ~ \int  {\textrm{d}\Omega_k \over 4 \pi} \, B^*_{ij}\, B_{lm} \, 
\left ( T^{ijlm} + T^{ijml} \right )\,, \nonumber 
\end{eqnarray}
where $B_{ij}\equiv  (1/2)(b_{ij}+b_{ji})$. 

Using the expression \eq{Tijlm} for $T^{ijlm}$, we  write 
\begin{eqnarray}
\label{R}
R (r, k) = R_1- R_2+R_3 
\end{eqnarray}
where
\begin{eqnarray}
& &R_1  =   2   \int {\textrm{d}\bar\Omega\over 4 \pi} ~ \int  {\textrm{d}\Omega_k \over 4 \pi} \,   \left( \sum_{i,j}~ B^*_{ij} B_{ij} 
\right)\,, \nonumber\\
& & R_2 =  4  \int {\textrm{d}\bar\Omega\over 4 \pi} ~ \int  {\textrm{d}\Omega_k \over 4 \pi} \, 
\left(\sum_{i,j} ~ B^*_{ij} B_{il} \frac{k_j k_l}{k^2} \right)\,, \nonumber \\
& & R_3 =   2 ~\int {\textrm{d}\bar\Omega\over 4 \pi} ~ \int  {\textrm{d}\Omega_k \over 4 \pi} \,  
\left(\sum_{i,j} B^*_{ij} B_{lm} \frac{k_i k_j k_l k_m}{k^4}  \right) \, .
\label{R12}
\end{eqnarray}

We assume isotropic turbulence, hence the $\vec k$  components satisfy
$$ \int \textrm{d}\Omega_k ~\frac{k_i k_j}{k^2}  = \delta_{ij} \int \textrm{d}\Omega_k ~  \frac{k^2_r}{k^2} \, ,$$
where $\delta_{ij}$ is the Kronecker symbol for $i,j=r,\theta,\phi$. 
As in \cite{Belkacem08a}, we then obtain
\begin{eqnarray}
R_1  &=&   2  \int {\textrm{d}\bar\Omega\over 4 \pi} ~    \left( \sum_{i,j}~ \vert B_{ij} \vert^2  \right)\,,
\nonumber
\\
R_2  &=&  2  \alpha ~R_1\,, \label{R2} \nonumber \\
R_3  &=&  \beta  ~R_1  + 2 ~\beta  ~  \left( \int {\textrm{d}\bar\Omega\over 4\pi} 
  ~\sum_{i \not=j} \left(B^*_{ii} B_{jj} + {\rm c.c.} \right)  \right)\,, 
\label{R3}
\end{eqnarray}
where we have set 
\begin{equation}
\alpha \equiv  \int {\textrm{d}\Omega_k \over 4 \pi} ~  \frac{k_r^2}{k^2} \quad\hbox{and}\quad
\beta \equiv \int {\textrm{d}\Omega_k \over 4 \pi} ~    \frac{k_r^4}{k^4}\,.
\end{equation} 

Using \eq{eqn:bij_4} to compute \eq{R}, with  $\alpha  = 1/3$ and $\beta  =  1/5$ \citep[see][for details]{Belkacem08a}, yields
\begin{eqnarray}
\label{Rfinal}
R (r) &=&   {16\over 15} ~  \left| \deriv{\xi_r}{r}  \right|^2  +  {44\over 15} ~    \left| \frac{\xi_r}{r}  \right|^2
+   \frac{4}{5} \left( \frac{\xi^*_r}{r} \deriv{\xi_r}{r} + {\rm c.c.} \right) \nonumber \\ 
&+&  ~ L^2 \left[ {11\over 15} ~ \left( \left| \mathcal{A}  \right|^2 + \left| \mathcal{B}  \right|^2 \right) - {22 \over 15} \left(\frac{\xi_r^* \xi_H}{r^2} + {\rm c.c.}\right) \right] \nonumber \\
&+&  \left| \frac{\xi_H}{r}  \right|^2 
\left( \frac{16}{15} L^4+ \frac{8}{5} {\cal F}_{\ell,\vert m \vert} - \frac{2}{3} L^2 \right) \nonumber \\
&+&  \left| \frac{\xi_T}{r}  \right|^2  \left( \frac{11}{5} L^2 (L^2-2) - \frac{8}{5} {\cal F}_{\ell,\vert m \vert} - \frac{2}{3} L^2 \right) \nonumber \\
 &-&{2\over 5} L^2 \left(\deriv{\xi^*_r}{r} {\xi_H \over r} 
+ {\rm c.c.} \right)\,,
\end{eqnarray}
where we have defined
\begin{equation}
\mathcal{A} =  \deriv{\xi_H}{r} +\frac{1}{r}(\xi_r-\xi_H) \quad\hbox{and}\quad \mathcal{B} =  \deriv{\xi_T}{r} - \frac{\xi_T}{r}\,\,
\end{equation}
while
\begin{equation}
{\cal F}_{\ell,|m|} =  \frac{|m| (2 \ell + 1)}{2} \left[ \ell\left(\ell+1\right) - (m^2 + 1 )\right]
\end{equation}
with $L^2 = \ell (\ell + 1)$.

\section{First-order perturbation of the excitation rates}
\label{first_order}

We recall the main results about the first-order perturbation of a spheroidal mode $\vec\xi^{\left(0\right)}$ due to the Coriolis acceleration (of frequency $\omega_{0}$) and establish the perturbation of the excitation rates. 
Following the classical method given in Unno et al. (1989), we obtain
\begin{equation}
\xi_{r}=\xi_{r;n,l}^{\left(0\right)}+\left(\frac{2\Omega}{\omega_0}\right)m\sum_{n^{'} \neq n}C_{n^{'},n,l}\;\xi_{r;n^{'},l}^{\left(0\right)}\, ,
\label{xirc}
\end{equation}
\begin{equation}
\xi_{H}=\xi_{H;n,l}^{\left(0\right)}+\left(\frac{2\Omega}{\omega_0}\right)m\sum_{n^{'} \neq n}C_{n^{'},n,l}\;\xi_{H;n^{'},l}^{\left(0\right)}
\label{xithc}
\end{equation}
and
\begin{equation}
\xi_{T}=i \left(\frac{2\Omega}{\omega_0}\right) z_{m}^{l}\, ,
\label{xiazc}
\end{equation}
where $\xi_{r;n^{'},l}^{\left(0\right)},\xi_{H;n^{'},l}^{\left(0\right)}$ are solutions of the oscillation equation without rotation \citep{Unno89},  
\begin{eqnarray}
\label{Cnl}
\lefteqn{C_{n^{'},n,l}=\frac{\omega_{0}^{2}}{\left(\omega_{0}^{2}-\omega_{0;n^{'}}^{2}\right)I_{n^{'}}}}\nonumber\\
& &\times\int_{0}^{R}\left[\xi_{H;n,l}^{\left(0\right)}\xi_{r;n^{'},l}^{\left(0\right)*}+\left(\xi_{r;n,l}^{\left(0\right)}+\xi_{H;n,l}^{\left(0\right)}\right)\xi_{H;n^{'},l}^{\left(0\right)*}\right]\rho_0 r^2{\rm d}r
\end{eqnarray}
and
\begin{eqnarray}
z_{l,m}&=&l D_{l,m}\left[\xi_{r;n,l-1}^{\left(0\right)}-\left(l-1\right)\xi_{H;n,l-1}^{\left(0\right)}\right]\nonumber\\
& &-\left(l+1\right)D_{l+1,m}\left[\xi_{r;n,l+1}^{\left(0\right)}+\left(l+2\right)\xi_{H;n,l+1}^{\left(0\right)}\right] \, ,
\end{eqnarray}
where 
\begin{equation}
D_{l,m}=\frac{1}{l^2}\sqrt{\frac{l^2-m^2}{4l^2-1}}.
\end{equation}
The Coriolis corrective terms are of the order of ${2\Omega}/{\omega_{{0}}}$, which is here assumed to be small. Inserting Eqs. (\ref{xirc}-\ref{xithc}-\ref{xiazc}) into Eq. (\ref{C2R_ref}), we obtain
\begin{equation}
C_{R,m}^{2}=\left[C_{R}^{\left(0\right)}\right]^2+ m \left(\frac{2\Omega}{\omega_{0}}\right)\left[C_{R,|m|}^{\left(1\right)}\right]^2
\end{equation}
where
\begin{equation}
\left[C_{R}^{\left(0\right)}\right]^{2}=4\pi^3\int{\rm d}m \; R^{\left(0\right)}S_{R}\left(\omega_{0}\right)\,,
\end{equation}
\begin{equation}
\left[C_{R,|m|}^{\left(1\right)}\right]^{2}=4\pi^3\int{\rm d}m \; R_{|m|}^{\left(1\right)} S_{R}\left(\omega_{0}\right)\, ,
\end{equation}
and $R^{\left(0\right)}$ and $R_{|m|}^{\left(1\right)}$ correspond to the perturbative expansion of $R$ given in Eq. (\ref{gammabeta}) where
\begin{eqnarray}
R^{\left(0\right)} &=&   {16\over 15} ~  \left| \deriv{\xi_{r;n,l}^{\left(0\right)}}{r}  \right|^2 \! + \! {44\over 15} ~    \left| \frac{\xi_{r;n,l}^{\left(0\right)}}{r}  \right|^2
\! + \!  \frac{4}{5} \left( \frac{\xi_{r;n,l}^{\left(0\right)*}}{r} \deriv{\xi_{r;n,l}^{\left(0\right)}}{r} + {\rm c.c.} \right) \nonumber \\ 
&+&  ~ L^2 \left[ {11\over 15} ~ \left( \left| \mathcal{A}  \right|^2 + \left| \mathcal{B}  \right|^2 \right) - {22 \over 15} \left(\frac{\xi_{r;n,l}^{\left(0\right)*} \xi_{H;n,l}^{\left(0\right)}}{r^2} + {\rm c.c.} \right) \right] \nonumber \\
&+&  \left| \frac{\xi_{H;n,l}^{\left(0\right)}}{r}  \right|^2 
\left( \frac{16}{15} L^4+ \frac{8}{5} {\cal F}_{\ell,\vert m \vert} - \frac{2}{3} L^2 \right) \nonumber \\
 &-&{2\over 5} L^2 \left(\deriv{\xi_{r;n,l}^{\left(0\right)*}}{r} {\xi_{H;n,l}^{\left(0\right)} \over r} 
+ {\rm c.c.} \right)\, ,
\end{eqnarray}
\begin{equation}
\label{R1}
R^{\left(1\right)}=\sum_{n' \neq n}C_{n^{'},n,l}\;f_{n^{'},n,l,|m|}
\end{equation}
where
\begin{eqnarray}
\label{perturb_cr}
f_{n^{'},n,l,|m|}&=&\frac{16}{15}\left[\frac{{\rm d}\xi_{r;n,l}^{\left(0\right)}}{{\rm d}r}\frac{{\rm d}\xi_{r;n^{'},l}^{\left(0\right)*}}{{\rm d}r}+{\rm c.c.}\right]\nonumber\\
&+&\frac{44}{15}\left[\frac{\xi_{r;n,l}^{\left(0\right)}\xi_{r;n^{'},l}^{\left(0\right)*}}{r}+{\rm c. c.}\right]\nonumber\\
&+&\left(\frac{16}{15}L^4-\frac{2}{3}L+\frac{8}{5}{\mathcal F}_{l,\vert m \vert}\right)\left[\frac{\xi_{H;n,l}^{\left(0\right)}\xi_{H;n^{'},l}^{\left(0\right)*}}{r^2}+{\rm c. c.}\right]\nonumber\\
&+&\frac{11}{15}L^2\Bigg[\left(\frac{{\rm d}\xi_{H;n,l}^{\left(0\right)}}{{\rm d}r}+\frac{\xi_{r;n,l}^{\left(0\right)}-\xi_{H;n,l}^{\left(0\right)}}{r^2}\right)\nonumber\\
& &\times\left(\frac{{\rm d}\xi_{H;n^{'},l}^{\left(0\right)*}}{{\rm d}r}+\frac{\xi_{r;n^{'},l}^{\left(0\right)*}-\xi_{H;n^{'},l}^{\left(0\right)*}}{r^2}\right)+{\rm c. c.}\Bigg]\nonumber\\
&+&\frac{4}{5}\left(\frac{{\rm d}\xi_{r;n,l}^{\left(0\right)}}{{\rm d}r}\frac{\xi_{r;n^{'},l}^{\left(0\right)*}}{r}+\frac{{\rm d}\xi_{r;n^{'},l}^{\left(0\right)}}{{\rm d}r}\frac{\xi_{r;n,l}^{\left(0\right)*}}{r}+{\rm c. c.}\right)\nonumber\\
&-&\frac{2}{5}L\left(\frac{\xi_{H;n,l}^{\left(0\right)}}{r}\frac{{\rm d}\xi_{r;n^{'},l}^{\left(0\right)*}}{{\rm d}r}+\frac{\xi_{H;n^{'},l}^{\left(0\right)}}{r}\frac{{\rm d}\xi_{r;n,l}^{\left(0\right)*}}{{\rm d}r}+{\rm c. c.}\right)\nonumber\\
&-&\frac{22}{15}L\left(\frac{\xi_{r;n,l}^{\left(0\right)}\xi_{H;n^{'},l}^{\left(0\right)*}}{r}+\frac{\xi_{r;n^{'},l}^{\left(0\right)}\xi_{H;n,l}^{\left(0\right)*}}{r}+{\rm c. c.}\right)\,.\nonumber\\
\end{eqnarray}
%and
%\begin{equation}
%R_{B}^{\left(1\right)}=\frac{32}{15}I\left(i\frac{\xi_{H;n,l,m}^{\left(0\right)}z_{l,m}^{*}}{r^2}+{\rm c. c.}\right).
%\end{equation}

\end{document}